%% file: main.tex
\documentclass{vgtc}                          

\graphicspath{{figures/}{pictures/}{images/}{./}} 

\usepackage{times}                     

\usepackage{tabu}                      
\usepackage{booktabs}                  
\usepackage{lipsum}                    
\usepackage{mwe}                       

\usepackage{mathptmx}                  

\vgtccategory{Research}

\vgtcinsertpkg

\title{\textit{VisAider}: AI-Assisted Context-Aware Visualization Support for Data Presentations}

\author{Kentaro Takahira\thanks{e-mail: ktakahira@connect.ust.hk}\\ %
        \scriptsize The Hong Kong University of Science and Technology %
\and Yuki Ueno\thanks{e-mail: yueno@asu.edu}\\ %
     \scriptsize Arizona State University 
     }
\input{0-abstract_wip}

\keywords{Data Storytelling, Data Presentation, AI-assisted Visualization}

\graphicspath{{figs/}{figures/}{pictures/}{images/}{./}} 

\usepackage{tabu}                      
\usepackage{booktabs}                  
\usepackage{lipsum}                    
\usepackage{mwe}                       
\usepackage{float} 

\usepackage{mathptmx}                  

\usepackage{enumerate}
\usepackage{enumitem}
\usepackage[svgnames]{xcolor}

\usepackage{booktabs} 
\usepackage{amssymb} 

\usepackage{booktabs} 
\usepackage{tabularx} 

\usepackage{xspace,xpunctuate}

\usepackage{tikz}  
\usetikzlibrary{shadings}
\definecolor{cback}{HTML}{E9ECEF}
\definecolor{cframe}{HTML}{495057}
\definecolor{clightblue}{HTML}{DDEFFC}
\definecolor{clightgreen}{HTML}{DFFFE5}
\definecolor{clightyellow}{HTML}{FFF4DB}

\definecolor{cback1}{HTML}{F8F9FA} 
\definecolor{cframe1}{HTML}{ADB5BD} 

\definecolor{cback2}{HTML}{F1F3F5} 
\definecolor{cframe2}{HTML}{CED4DA} 

\definecolor{cback3}{HTML}{EBF5FB} 
\definecolor{cframe3}{HTML}{A9CCE3} 

\definecolor{cback4}{HTML}{FDF3E7} 
\definecolor{cframe4}{HTML}{F5CBA7} 

\newcommand*\circled[1]{\tikz[baseline=(char.base)]{
    \node[shape=rectangle,rounded corners=1.5pt,fill=cback,text=black,draw=cframe,inner sep=1pt] (char) {#1};}}

\begin{document}



\input{1-introduction_wip}
\input{2-related_works_wip}
\input{3-formative-study}
\input{4-VisAider}
\input{5-conclusion}

\bibliographystyle{abbrv-doi}
\bibliography{reference}

\end{document}

%% file: 0-abstract_wip.tex
\abstract{
Effective real-time data presentation is essential in small-group interactive contexts, where discussions evolve dynamically and presenters must adapt visualizations to shifting audience interests. However, most existing interactive visualization systems rely on fixed mappings between user actions and visualization commands, limiting their ability to support richer operations such as changing visualization types, adjusting data transformations, or incorporating additional datasets on the fly during live presentations. This work-in-progress paper presents \textit{VisAider}, an AI-assisted interactive data presentation prototype that continuously analyzes the live presentation context, including the available dataset, active visualization, ongoing conversation, and audience profile, to generate ranked suggestions for relevant visualization aids. Grounded in a formative study with experienced data analysts, we identified key challenges in adapting visual content in real time and distilled design considerations to guide system development. A prototype implementation demonstrates the feasibility of this approach in simulated scenarios, and preliminary testing highlights challenges in inferring appropriate data transformations, resolving ambiguous visualization tasks, and achieving low-latency responsiveness. Ongoing work focuses on addressing these limitations, integrating the system into presentation environments, and preparing a summative user study to evaluate usability and communicative impact.
}

\keywords{Immersive Analytics, Egocentric Network, Network Visualization, Virtual Reality}

\teaser{
  \centering
\includegraphics[width=1\linewidth]{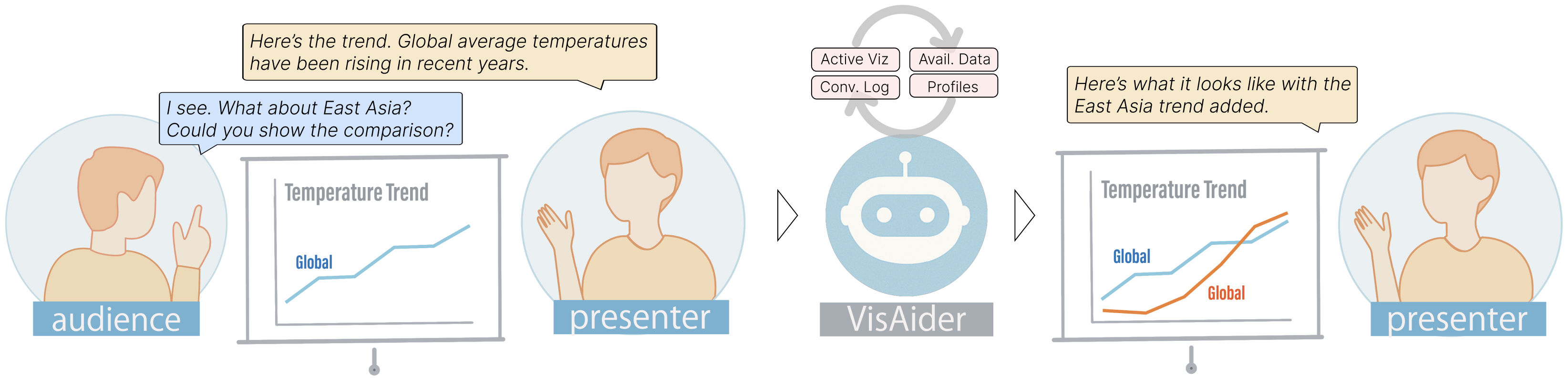}
  \caption{%
\textit{VisAider}: An early-stage prototype with an AI agent that suggests visualization aids in real time based on the available dataset, active visualization, ongoing conversation, and audience profile. The system is designed to support more improvisational interactions for presenters during live presentations and is currently under development. }
  \label{fig:teaser}
}

%% file: 1-introduction_wip.tex
\firstsection{Introduction}
\maketitle
Effective real-time data presentation is essential in interactive, small-scale settings where dynamic dialogue between presenters and audiences frequently shapes the narrative~\cite{FromJamtoRecital}. 
In such situations, presenters must adapt visual content on the fly, tailoring data communication to audience interests, expertise, and emerging questions~\cite{howDataScienceIntermediate, FromJamtoRecital}. 
Existing interactive presentation systems support real-time manipulation of visualizations through modalities such as gestures, speech, and tangible objects~\cite{chironomia, jollygesture, tangibleNet, LeeSketchStory, realityTalk, insitutale, BodyDrienGraphics}, but they typically rely on static, predefined mappings between user actions and visualization commands.
While effective for basic operations (e.g., highlighting points, switching views), these mappings often fall short when richer expressiveness is needed, such as changing visualization types, adjusting data transformations, or incorporating additional datasets, in response to evolving discussions.

Preparing every possible visualization in advance is impractical, and creating them spontaneously can disrupt the presentation flow.
To address this challenge, this work-in-progress paper presents \textit{VisAider}, an AI-assisted presentation tool with an agent that continuously analyzes the live presentation context, including 1) the available dataset, 2) the currently presented visualization, 3) the ongoing conversation, and 4) the audience profile, to generate ranked suggestions for relevant visualization aids.
Presenters can incorporate these suggestions through lightweight interactions, allowing flexible, real-time adaptation without interrupting the narrative flow.

Our approach is informed by a formative study with three experienced data analysts who regularly deliver live data presentations. 
The study identified key challenges in adapting visual content during a presentation and resulted in four design considerations: 1) preserving presenter autonomy, 2) enabling low-effort integration, 3) ensuring clear relevance to the presented content, and 4) adapting the format of visualizations to the audience's profile.
Guided by these findings, we developed \textit{VisAider}, a prototype that employs a multi-stage pipeline based on Large Language Models (LLMs) for contextual analysis, data selection, candidate generation, candidate evaluation, and Vega-Lite specification creation.

Preliminary tests in synthetic presentation scenarios show that \textit{VisAider} can select relevant datasets, identify appropriate columns, and focus on data ranges that match discussion cues. However, the system currently faces four main limitations: 1) difficulty inferring complex data transformations from conversational context, 2) occasional ambiguity in determining the intended visualization task, 3) latency of 20–30 seconds that disrupts presentation flow, and 4) limited integration with existing presentation tools. 
These issues guide our ongoing improvements, which include constraining the scope of data transformations, leveraging longer conversation histories, streamlining processing to achieve faster generation, and embedding the tool within common presentation environments.

For future work, we aim to complete a fully functional prototype, conduct a summative user study in real presentation settings, and refine our design framework to inform next-generation AI-assisted presentation systems that enable fluid, context-aware, and audience-responsive data communication.

%% file: 2-related_works_wip.tex
\section{Related Works}
Our research centers on synchronous data storytelling, AI-supported communication, and AI-driven visualization.

\subsection{Synchronous Data Storytelling}
Real-time data presentation is particularly important in small-group and interactive settings~\cite{howDataScienceIntermediate, FromJamtoRecital}. 
Prior systems have enabled such interaction through multimodal inputs—gestures, speech, or tangible objects~\cite{chironomia, jollygesture, tangibleNet, LeeSketchStory, realityTalk, insitutale, BodyDrienGraphics, VisConductor}.  
These works have explored making interactions intuitive for presenters, preserving presentation flow, and increasing expressive power to support diverse visualization commands.
However, most rely on predefined mappings between user actions and interaction commands (e.g., filtering or selecting data points) ~\cite{chironomia, VisConductor, tangibleNet, insitutale}. 
Such rigid coupling restricts the presenter's ability to generate or adapt visualizations on the fly, particularly when responding to spontaneous audience questions or shifts in the discussion's focus.  
Our work differs in that it introduces an AI-driven agent that continuously recommends relevant visualization aids during a presentation, enabling expressive data communication that evolves with the interaction between the audience and the presenter, while preserving the presenter's control over which suggestions to incorporate.

\subsection{AI-Supported Communication}
AI agents are increasingly used to augment human communication, particularly in scenarios involving real-time conversational interpretation and retrieval of relevant materials~\cite{AreWeOnTheRight, crosstalk, realityChat, VisualCaptions, ExploringPotentialOfAI, TheKnownStranger}. 
These systems typically analyze dialogue context to surface pertinent information and integrate it into the communication process~\cite{realityChat, crosstalk}.  
However, most prior work focuses on general-purpose conversation support rather than addressing the domain-specific challenges of interactive, data-rich presentations. 
Such presentations require interpreting data communication intent, considering the available dataset, inferring appropriate visualization tasks, and adapting to evolving audience profiles. 
Our research advances this space by designing an AI agent specifically tailored to the unique demands of live data presentations.

\subsection{LLMs for Visualization Generation}
Recent advances in LLMs have spurred a growing body of research on automated visualization generation~\cite{areLLM, DynaVis, lida, chartGPT}.
Most existing systems focus on producing executable visualization code, such as Vega-Lite specifications~\cite{DynaVis, lida, chartGPT}, which supports validation, reproducibility, and post-generation user modification.
A common approach is to design prompts that include example schemas with placeholders, enabling the LLM to infer and fill the missing elements via few-shot learning before the results are validated~\cite{DynaVis}.
Although effective in offline or asynchronous scenarios, these methods typically depend on explicit user prompts describing the desired visualization. 
Such reliance can be impractical in live presentations, where maintaining narrative flow is critical and manual prompt engineering is disruptive.
Our work builds on this foundation by investigating LLM-based visualization generation in dynamic presentation contexts, where the system must adapt in real time to factors such as audience profiles, evolving discussion topics, and shifting focal points, without interrupting the presenter's delivery.

%% file: 3-formative-study.tex
\section{Formative Study}
We conducted semi-structured interviews with three experienced data analysts (P1–P3), each with more than eight years of professional experience, who regularly deliver live presentations in domains including retail, media, and manufacturing.
The goal was to identify current challenges in delivering data presentations, explore opportunities for real-time visualization support, and distill design considerations for AI-assisted presentation tools. Each 60-minute session was documented in detail through interviewer notes.

\subsection{Current Challenges in Presentations}
All participants reported frequently needing to adapt their content during small-group meetings in response to audience reactions. Three primarily used slide decks, two used spreadsheets (Excel), one used Confluence, one used Streamlit, and one used an in-house dashboard. 
P1 explained a preference for bringing datasets in an adjustable format rather than static images: \textit{``If I think the discussion might go in directions I can’t predict, I bring Excel or a dashboard so I can adjust the data on the spot and respond to what the client is asking.''} 
P2 and P3 described taking a similar approach, favoring tools that allow on-the-spot adjustments to meet audience requests.

Participants described common audience-driven adaptation requests, including adding new data, changing the displayed data range, switching chart types, and modifying aggregation granularity.
P1 noted that such requests are typically related to the presented content and highlighted the importance of making these relationships explicit: \textit{``In retail data, there are multiple levels of granularity, such as product categories and SKUs (Stock Keeping Units), and many questions are interrelated. If the system could clearly show these relationships, it would support the story much better.''} P2 highlighted the role of AI in client communication, noting their in-house dashboard supports natural language queries, enabling flexible client communication and fostering more interactive exchanges between presenter and audience.

Participants also emphasized that the preferred presentation format often depends on the audience profile.
As P2 explained: \textit{``An overview works best with charts, but if the audience is highly data-literate or deeply familiar with the topic, I sometimes present raw numbers instead.''} 
P1 added that matching the level of detail to the audience’s expertise helps maintain engagement.
From these discussions, participants agreed that effective visualization aids should take into account three factors: 1) the available datasets, 2) the visualization currently being presented, and 3) the audience’s interests.

\subsection{Design Considerations}
From these results, we distilled four key design considerations:

\noindent\textbf{\circled{DC1} Preserve Presenter Autonomy:}  
While participants welcomed AI-generated visualization suggestions, they stressed that presenters must retain control over which aids to adopt. As P1 noted: \textit{``Even if the audience asks for it, there are cases where I don’t want to show certain information. That decision should be the presenter’s.''}

\noindent\textbf{\circled{DC2} Enable Low-Effort, Rapid Integration:}  
Visualization aids should be quick and easy to incorporate without interrupting the presentation flow, a need also highlighted in prior work~\cite{tangibleNet}.

\noindent\textbf{\circled{DC3} Ensure Clear Relevance to Active Content:}  
Suggested visualizations should be explicitly tied to the presenter’s prepared content and easily interpretable in that context. They should clearly communicate their connection to the ongoing presentation.

\noindent\textbf{\circled{DC4} Adapt Format to Audience Profile:}
Visualization aids should be presented in a format suited to the audience's level of expertise, familiarity with the topic, and specific interests, ensuring both clarity and engagement.

%% file: 4-VisAider.tex
\begin{figure*}[t]
    \centering
    \includegraphics[width=1\linewidth]{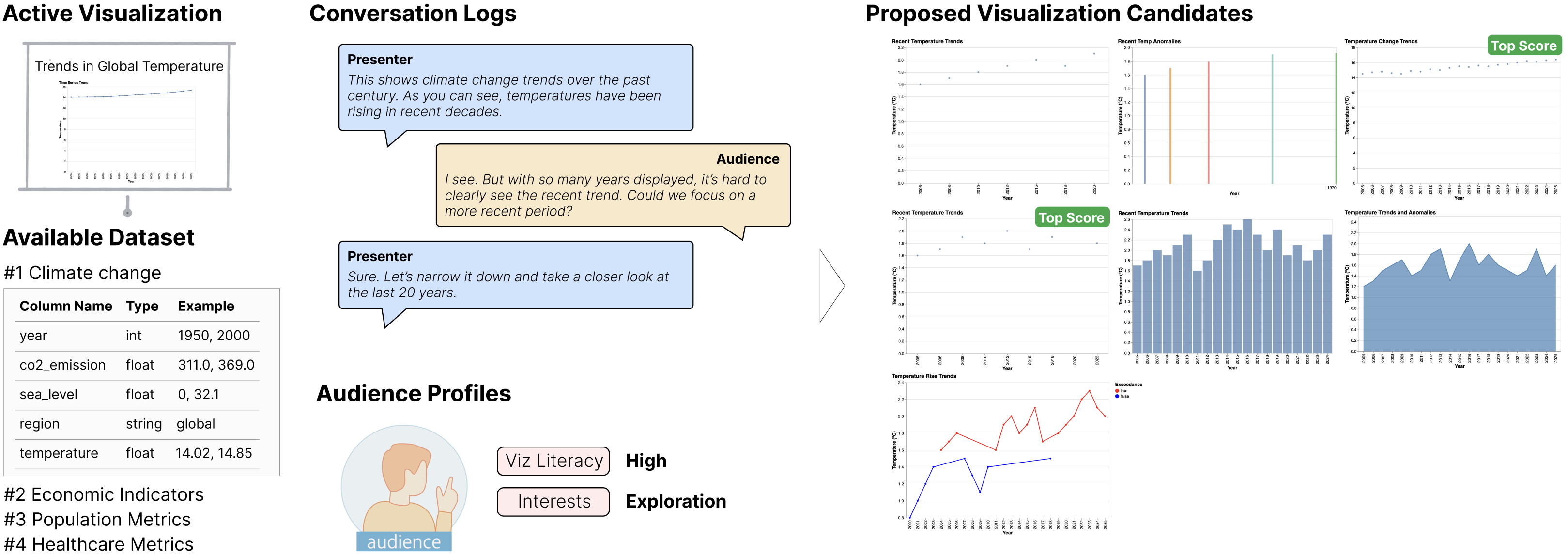}
     \caption{Example scenario on trends in global temperature, showing active charts, available dataset, presenter–audience conversation, audience profile, and the system's inferred visualization candidates.}
    \label{fig:example_result}
\end{figure*}

\section{Prototyping}
Guided by the formative findings, we developed \textit{VisAider}, a prototype designed to explore feasibility and stimulate further discussion on design trade-offs.

\subsection{Prototype Design}
\textit{VisAider} was designed to address four central needs: 1) preserving presenter autonomy, 2) enabling low-effort, real-time integration, 3) ensuring clear relevance to the active content, and 4) adapting the format to the audience profile. To operationalize these considerations, the system takes into account four key factors when determining appropriate visualization aids during a presentation: 1) the available dataset, 2) the presented content, 3) the ongoing conversation, and 4) the audience profile. Based on these factors, the system generates candidate visualization aids (\cref{fig:example_result}).

\begin{figure}[H]
    \centering
    \includegraphics[width=1\linewidth]{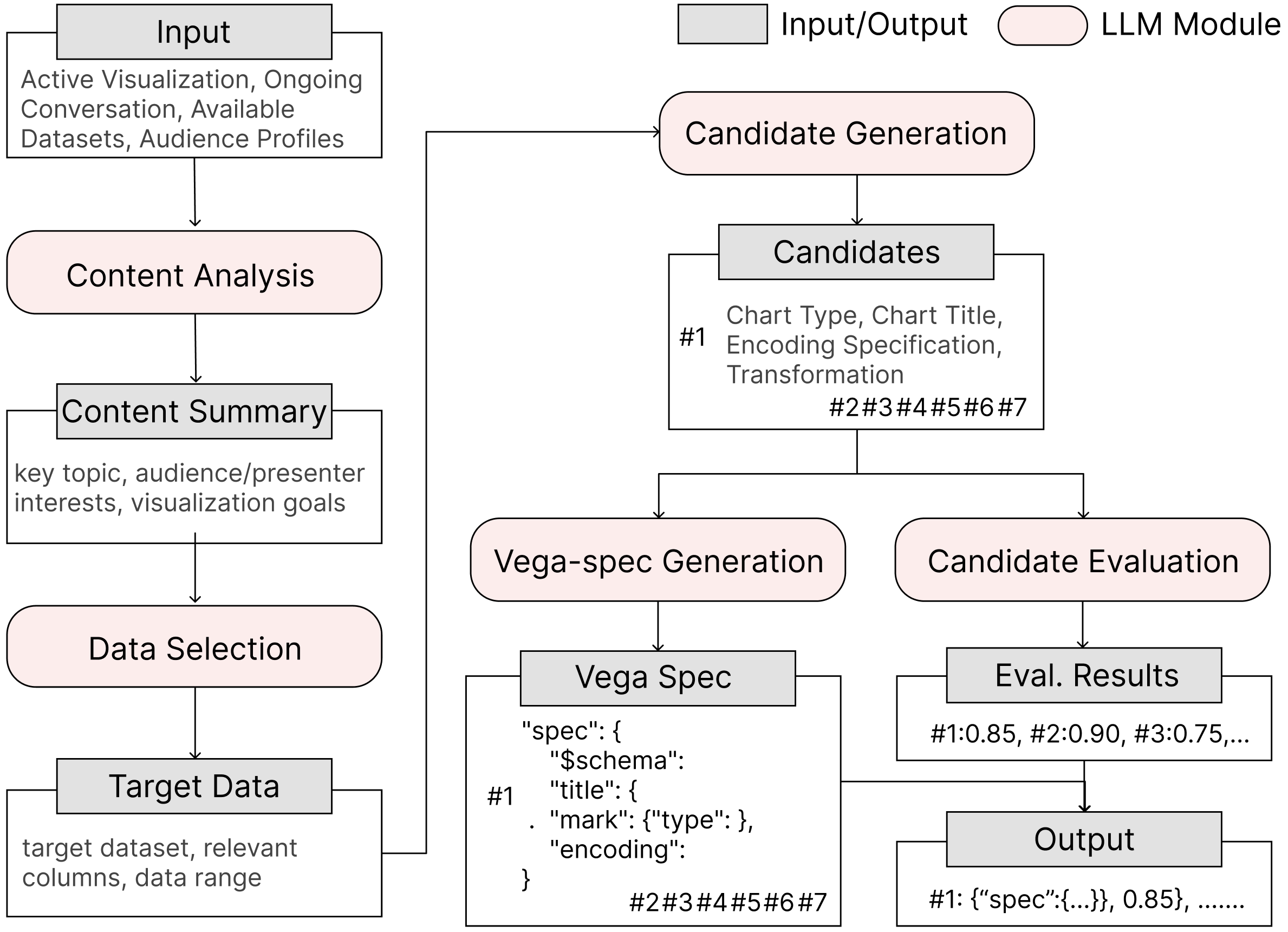}
     \caption{The \textit{VisAider} pipeline for generating audience-responsive visualization aids.}
    \label{fig:pipeline}
\end{figure}

\textit{VisAider}'s AI-agent comprises five modules--\textit{content analysis}, \textit{data selection}, \textit{candidate generation}, \textit{Vega specification generation}, and \textit{candidate evaluation}, orchestrated using LangChain\footnote{\url{https://www.langchain.com}}, with all LLM-based components powered by GPT-4o\footnote{\url{https://openai.com/index/hello-gpt-4o/}}~(\cref{fig:pipeline}). 
In the \textbf{content analysis} module, the agent analyzes the conversation transcript, the dataset in use, and the currently displayed visualization. From this, it infers 1) the main topic under discussion, 2) the key points being addressed, 3) the audience's interests, and 4) the objectives that visualizations should achieve in the current context. 
These inferences, combined with the available dataset, are then passed to the \textbf{data selection} module, which determines the target dataset, relevant columns, and appropriate data ranges for visualization.

In the \textbf{candidate generation} module, the agent produces a set of visualization aid candidates, each defined by a chart type, chart title, encoding specification, and any applicable data transformations. 
To promote diversity, the prompts explicitly instruct the LLM to generate a range of visualization types and perspectives. 
These candidates are then processed in parallel by the \textbf{candidate evaluation} module, which scores them based on how well they align with the inferred presentation needs, and the \textbf{Vega specification generation} module, which uses few-shot prompting with examples of line charts, pie charts, bar charts, and scatter plots to produce valid, coherent Vega-Lite specifications. By executing evaluation and specification generation in parallel, the system reduces processing time, helping to maintain responsiveness suitable for real-time presentation scenarios. Finally the ranked list of visualizations is produced by ordering the candidates according to their evaluation scores. 

\noindent

\subsection{Preliminary Result}
Early tests in synthetic presentation scenarios suggest that the system can interpret key aspects of the presentation context. 
In several trials, it correctly identified the most relevant dataset from multiple options, selected appropriate columns, and constrained the data range to match the inferred discussion focus. 
For example, given a dataset of global climate change records from 1950 to 2025 and a conversation in which an audience member found it difficult to see recent trends, prompting the presenter to focus on ``the most recent 20 years'', the system narrowed the range to roughly the last 20 years and generated candidate charts such as a line chart highlighting recent temperature changes and a bar chart comparing average annual values (\cref{fig:example_result}).
These results demonstrate the feasibility of generating visualization aids dynamically without explicit prompting, aligning with \circled{DC2} by reducing presenter interaction effort.

The tests also exposed important limitations. 
The system struggled with complex data transformations implied by the conversation, for instance, calculating year-over-year change rates from raw values, resulting in incomplete or inaccurate visualizations. 
It sometimes misidentified the intended visualization task when conversational cues were ambiguous, producing only tangentially relevant candidates. 
Latency was another major challenge: generating eight ranked candidates required 20–30 seconds, interrupting the presentation flow and violating design requirements identified in prior studies.
While the pipeline can produce contextually relevant visualizations, these findings highlight the need for better transformation inference, task detection, and responsiveness before it can reliably support real-time use.


\subsection{Addressing Current Limitations}
The preliminary findings point to four main areas for improvement. 
First, although the system can select appropriate datasets, columns, and ranges, it struggles to infer and apply correct data transformations from the conversation. 
We plan to limit transformation types to a predefined, well-tested set of operations to improve stability and predictability in live settings.
Second, the system occasionally fails to detect the intended visualization task due to limited reasoning or insufficient context. 
We aim to make task resolution easier by leveraging longer conversation histories, referencing a broader range of presentation content, and presenting candidates with concise descriptions and context cues so presenters can quickly identify and adopt the most relevant option.
Third, latency remains high at 20–30 seconds for eight candidates, disrupting presentation flow. 
We plan to optimize module execution and parallelization and introduce caching strategies, with the goal of reducing generation time to a level that supports seamless real-time interaction.
Finally, the current standalone web interface limits integration into real presentation workflows. 
To address this, we will embed the system into common presentation tools (e.g., slideware, notebook) and streamline lightweight interaction methods so presenters can accept or reject suggestions without interrupting their delivery.

%% file: 5-conclusion.tex
\section{Conclusion}
We presented \textit{VisAider}, an AI-assisted interactive data presentation prototype designed to support real-time, audience-responsive visualization generation.
From a formative study with professional data analysts, we identified key challenges in adapting visual content to evolving audience needs and distilled four design considerations: 1)
preserving presenter autonomy, 2) enabling low-effort integration,
3) ensuring clear relevance to the presented content, and 4) adapting the format of visualizations to the audience's profile.
The current system processes four key inputs: the available dataset, the active visualization, the ongoing conversation, and the audience profile, to generate ranked visualization candidates. Preliminary testing demonstrates the feasibility of this approach but also reveals challenges, including difficulty inferring appropriate data transformations from conversational cues, ambiguity in determining the intended visualization task, and latency that limits responsiveness.
Our ongoing work focuses on improving inference accuracy, reducing latency for fluid interaction, and integrating the system with presentation tools for seamless practical use. Future work will complete a fully functional prototype, conduct a summative user study to assess usability and communicative impact, and contribute design insights for next-generation AI-driven presentation systems.